# Anisotropic electron-phonon coupling in the spinel oxide superconductor

Ge He[1], Yanli Jia[1], Xingyuan Hou[1], Zhongxu Wei[1], Haidong Xie[1], Zhenzhong Yang[1], Jinan Shi[1], Jie Yuan[1], Lei Shan[1,3], Beiyi Zhu[1], Hong Li[1], Lin Gu[1,3,4], Kai Liu[2,*], Tao Xiang[1,3], and Kui Jin[1,3,*]

[1]Beijing National Laboratory for Condensed Matter Physics, Institute of Physics, Chinese Academy of Sciences, Beijing 100190, China.
[2]Beijing Key Laboratory of Opto-electronic Functional Materials & Micro-nano Devices, Department of Physics, Renmin University of China, Beijing 100872, China.
[3]Collaborative Innovation Center of Quantum Matter, Beijing, 100190, China.
[4]School of Physical Sciences, University of Chinese Academy of Sciences, Beijing 100190, China.

Among hundreds of spinel oxides, $LiTi_2O_4$ (LTO) is the only one that exhibits superconductivity ($T_c$ ~13 K). Although the general electron-phonon coupling is still the main mechanism for electron pairing in LTO, unconventional behaviors such as the anomalous magnetoresistance, anisotropic orbital/spin susceptibilities, etc. reveal that both the spin and the orbital interactions should also be considered for understanding the superconductivity. Here, we investigate tunneling spectra of [111]-, [110]- and [001]-oriented high quality LTO thin films. Several bosonic modes in tunneling spectra are observed in the [111]- and [110]-oriented films but not in [001]-oriented ones, and these modes still exist at $T \approx 2T_c$ and beyond the upper critical field, which are confirmed as stemming from electron-phonon interaction by DFT calculations. These modes only appear in special surface orientations, indicating that the electron-phonon coupling in LTO system is highly anisotropic and may be enhanced by orbital-related state. The anisotropic electron-phonon coupling should be taken seriously in understanding the nature of LTO superconductivity.

## I. INTRODUCTION

Oxides with spinel structure have attracted a broad attention for their extraordinary features, such as spin fluctuations from frustrated magnetic sublattice [1], charge ordering by mixed valence [2], and orbital ordering [3]. Accordingly, these intriguing properties generate rich functionalities including ferrimagnetism, magnetostriction, multiferroicity, etc [4-6]. So far, only a few of spinels show superconductivity, among which $LiTi_2O_4$ (LTO) is the only oxide and holds the highest record of superconducting critical transition temperature, i.e. $T_c$ ~ 13 K [7]. Therefore, it is interesting to study the nature of such considerable superconductivity, especially for a complicated system with frustrated Ti-sublattice and equal numbers of $Ti^{3+}$ and $Ti^{4+}$ ions.

The superconductivity of LTO was discovered by Jahnston *et al.* in 1973 [8]. It is commonly accepted to be an *s*-wave BCS superconductor with an intermediate electron-phonon coupling in previous studies [9,10]. However, an enhanced coupling constant ($\lambda_{tot}$ ~ 1.53) has been revealed from magnetic susceptibility measurements [11], compared to the phonon-assisted electronic correlations (i.e. $\lambda_{e-ph}$ ~ 0.65) [12]. This means that *d-d* electron correlations involving spin-orbit fluctuations cannot be ignored, which is evident from resonant inelastic soft-x-ray scattering (RIXS) [13] and nuclear magnetic resonance (NMR) measurements [14]. However, the lack of high quality single crystals prevents further exploration on the nature of its superconductivity.

Most recently, high quality LTO single crystal thin films were successfully synthesized by pulsed laser deposition (PLD) [15-17]. A careful study of electrical transport and tunneling spectra has been carried out on [001]-oriented LTO thin films, yielding the first electronic phase diagram against temperature and magnetic field [18]. In that work, experimental results point to an orbital-related state below ~ 50 K as well as remarkable spin fluctuations at higher temperature. The existence of spin-orbit fluctuations, which is crucial to high $T_c$ in cuprates [19] and Fe-based superconductors [20], thereby makes this system more intriguing. Nevertheless, general electron-phonon coupling is still a main candidate for the forming of Cooper pairs [12,21]. Prior to clarifying the electron pairing medium in LTO, the vital issue is to provide comprehensive experimental information on the electronic structure. Obviously, probing the information of Fermi surface and possible anisotropic correlations relies on high quality thin films of different orientations in this system with such a three dimensional lattice structure.

In this work, we present and compare the tunneling spectra of LTO single crystalline thin films deposited on [111]-, [110]- and [001]-oriented $MgAl_2O_4$ substrates by point contact spectroscopy. Remarkable bosonic modes are identified from the differential conductance of both the [111]- and [110]-oriented samples. These modes can persist from the superconducting state into normal state, surviving beyond 28 K and up to 16 T. Interestingly, they never show up in the [001]-oriented samples. We confirm that the modes appearing in some special surface orientations come from an intrinsic anisotropic electron-phonon coupling in LTO, which should be taken into account in understanding the unique superconductivity of this system.

## II. TUNNELING EXPERIMENTS

The [111]-, [110]-, and [001]-oriented LTO thin films were deposited on [111]-, [110]- and [001]-oriented $MgAl_2O_4$ substrates by PLD technique, respectively. All these epitaxial films used in this study show $T_c$ of ~ 11 K, and sharp transition of less than 0.5 K [22]. Point-contact measurements were performed by employing a home-made probe compatible with commercial cryostat that allows temperature down to 2 K and field up to 16 T. The sample stage with high precision of less than 0.1 µm, can move in range of 3 mm and 2.5 mm along the *z-* and *x-axes*, respectively. We use Pt-Ir tips to guarantee steady point contact junctions along the out-of-plane direction of the samples. All the junction resistances are about 50 ~ 60 Ω in our measurements, which are in the Sharvin limit, where the point contacts are in the ballistic regime, i.e. point contact radius is smaller than both the mean free path and the superconducting coherence length [23].

FIG. 1 displays the temperature dependence of differential conductance on samples of different orientations. The superconducting coherence peaks become weak with increasing the temperature and disappear at $T_c$, which naturally rules out heating effects in our measurements. For [001]-oriented samples, the tunneling spectra are almost the same as our previous report [18]. Interestingly, the spectra on [110]- and [111]-oriented samples are quite distinct from the [001]-oriented ones, where prominent humps are observed at bias voltages of ± 15 mV ($h_1$) and ± 35 mV ($h_2$) in the superconducting state, and can extend to more than 28 K. When the applied magnetic field parallels to the tunneling direction, the superconducting coherence peaks are suppressed and disappear at the upper critical field $H_{c2}$, whereas the humps $h_1$ and $h_2$ are almost unchanged under the field.

In order to catch the details of the humps, the spectra of [110]- and [111]-oriented samples in the normal state are compared with that in the superconducting state. As seen in FIG. 2, for both the [110]- and the [111]-oriented samples, the amplitude of $h_2$ remains almost unchanged with increasing temperature or field, whereas it is difficult to judge the change in $h_1$ due to the long tail of the coherence peaks.

To obtain the superconducting energy gap, we use a modified Blonder-Tinkham-Klapwijk (BTK) model to fit the normalized tunneling spectra in the superconducting state. In the BTK model, the tunneling regime is achieved for $Z > 1$, where $Z$ represents the tunneling barrier height and the Fermi velocity mismatch [24]. The modified BTK model uses a complex energy $E' = E + i\Gamma$, where the imaginary part ($\Gamma$) takes into account a finite quasiparticle lifetime by scattering [25]. The fitting parameters $Z$ and $\Gamma$ should not depend on the temperature and the pair breaking effect induced by magnetic field is akin to enhancing $\Gamma$ [26].

FIG. 3 shows the normalized tunneling spectra as well as their fittings for the [110]- and [111]-oriented films versus the temperature (see FIG. 3(a) and 3(b)) and the field (see FIG. 3(c) and 3(d)). The procedures of normalization and fitting have been described in details in Ref. [18]. $Z$ is about 2.1~2.3, confirming that our measurements are in the tunneling limit. The temperature and field dependences of superconducting energy gap Δ are shown in FIG. 3(e) and 3(f), respectively. First, Δ($T$) agrees with the BCS theory for all the samples of different orientations, and $2\Delta/k_BT_c$ is ~ 4. This value is roughly consistent with the results from specific heat [9] and Andreev reflection experiments [10] on LTO polycrystals. Second, the relation of Δ($H$) ~ -$H^2$, an indication of orbital-related state found in [001]-oriented films [18], also works for our [110]- and [111]-oriented samples.

## III. BOSONIC MODES ORIGINATION

In tunneling experiments, the features in the differential conductance at high bias often result from electron-boson interactions, e.g. phonons [27,28], spin resonance [29-31], etc. To catch the fine structures of the tunneling spectra, we plot the second derivative of the tunneling current curves, i.e. $d^2I/dV^2$. In FIG. 4(b), the $d^2I/dV^2$ curve of the [111]-oriented film exhibits two peaks (-17.6 ± 1 meV and -40 ± 1 meV) and two dips (-13.2 ± 1 meV and -28.8 ± 1 meV) at the negative bias, whereas two corresponding dips (16.4 ± 1 meV and 38.2 ± 1 meV) and two corresponding peaks (14.1 ± 1 meV and 28.5 ± 1 meV) show up at the positive bias, respectively. These features are verified in the [110]-oriented samples as well. These peaks/dips survive in the normal state up to 28 ± 2 K and 16 T (see FIG. 4(c) and 4(d)).

In unconventional high-$T_c$ superconductors, the bosonic modes from spin resonance disappear at $T_c$ such as in $Pr_{0.88}LaCe_{0.12}CuO_4$ [29], $Pr_{2-x}Ce_xCuO_{4-\delta}$ [32], $Ba(Fe_{1-x}Co_x)_2As_2$ [33], LiFeAs [34], and $Ba_{0.6}K_{0.4}Fe_2As_2$ [31]. Moreover, these modes should satisfy $\Omega/2\Delta < 1$ [35,36], where $\Omega$ is the bosonic mode energy. Our observations do not follow these features, thus rule out the possibility of spin-derived bosonic modes. Alternatively, the bosonic modes may originate from phonons. As seen in FIG. 4(b), the bias positions of the peaks and dips are close to the energy of phonon modes from our following density functional theory calculations as well as the phonon spectra of LTO polycrystals probed by neutron inelastic scattering [37].

In order to clarify the possible origin of these bosonic modes, we have studied the phonon spectra and electronic structures of LTO by density functional theory (DFT) calculations. LTO belongs to the space group of *Fd3m*, where the lithium and titanium cations are located at the tetrahedral 8a and octahedral 16d sites, respectively [11]. For the primitive cell of bulk LTO containing 14 atoms, there are 39 optical phonon modes at the Brillouin zone center. We find four phonon modes, i.e. the $T_{2u}(2)$, $E_u(2)$, $T_{1u}(4)$ and $E_g$ modes, with energy below 42 meV (see TAB. A1). The $T_{2u}(2)$ and $E_u(2)$ modes come from the Ti atom vibrations. The relative movement of Ti and Li atoms contributes to the $T_{1u}(4)$ mode. The $E_g$ mode is attributed to the O atom vibration. Among them, the calculated frequencies of the $T_{2u}(2)$ mode (14.5 meV), the $E_u(2)$ mode (25.3 meV) and the $E_g$ mode (41.1 meV) coincide with the peak/dip positions observed in the $d^2I/dV^2$ spectra.

## IV. DISCUSSION AND CONCLUSION

Our new findings can be summarized as follows: First, bosonic modes are observed in the tunneling spectra of the [110]- and [111]-oriented films, except for the [001]-oriented ones. Second, the bosonic modes that are discernable in the superconducting state still exist at $T = 2T_c$ and beyond the $H_{c2}$. The energies of these modes are consistent with those of phonons modes confirmed by our DFT calculations and the neutron inelastic scattering experiments. Next, we will focus on two fundamental issues, i.e. the anisotropic spectra and the robustness of these bosonic modes.

The primary concern is the absence of humps in the tunneling spectra of the [001]-oriented samples (see FIG. 1). We first study the influence of substrate orientation on the electronic structures. Taking into account the lattice distortion from the strain of the substrate, the band structures of the [111]-oriented film (solid lines in FIG. 5(a)) and the [001]-oriented film (solid lines in FIG. 5(b)) vary from that of the ideal bulk crystal (solid lines in FIG. 5(c)). When a typical phonon mode (the 41.1 meV $E_g$ mode) is excited, the band structures show dramatic changes for the [111]-oriented film (dash lines in FIG. 5(a)) but not for the [001]-oriented film (dot lines in FIG. 5(b)), indicating the relatively weak electron-phonon coupling for the bands around the Fermi level with the $E_g$ phonon mode in the [001]-oriented film. Thus, one explanation to the anisotropic spectra is the consequence of lattice distortions from substrates, which is extrinsic to bulk LTO. The above DFT calculations consider a 2% lattice distortion for the [001]-oriented film. This corresponds to ~ 0.16 Å reduction of the in-plane lattice constants, which is at least one order of magnitude larger than the measured value (~ 0.01 Å) from X-ray diffraction. Moreover, the substrate strain should be released as the thickness of our samples are ~ 200 nm.

Alternatively, the anisotropic spectra are intrinsic to bulk LTO. The Li/Ti ions occupy the tetrahedral/octahedral sites and the O atoms displace from their ideal positions [21], thus Jahn-Teller distortions are expected to exist and take effect. In this case, the Ti-O octahedrons, which allow a tilted angle of several degrees [38], result in anisotropic modulations to the ideal lattice structure, and accordingly account for the anisotropic spectra as tunneling into different surfaces of LTO. We perform further DFT calculations to inspect our conjecture. For a typical phonon mode in the bulk LTO, the $E_g$ phonon mode (~ 41.1 meV) consists of O vibrations around Ti atoms (see FIG. 5(d)). Once the internal atomic positions of bulk LTO are displaced according to the $E_g$ mode, the enhanced Jahn-Teller distortions of $TiO_6$ octahedrons alter the bands around the Fermi level distinctly along different directions of Brillouin zone (short dash lines in FIG. 5(c)), i.e. an anisotropic electron-phonon coupling. As long as the oxygen displacements induced by the excitation of this $E_g$ mode are strong enough, the number of bands crossing the Fermi level may change along the Γ-L direction but not along the Γ-X direction. Considering the conductance of the [001]-oriented film is related to the bands along the Γ-X direction, the invariant bands crossing the Fermi level may result in the absence of humps in the tunneling spectra. Furthermore, the Cs-corrected annular-bright-field (ABF) scanning transmission electron microscopy (STEM) directly observed ordering of oxygen vacancies in some domains of LTO films as shown in FIG. 5(e) and 5(f). These ordered O vacancies around Ti atoms lead to strong Jahn-Teller distortions as seen in FIG. 5(f). In this case, the oxygen-vacancy enhanced Jahn-Teller distortions may associate with the anisotropic electron-phonon coupling of the bulk LTO and possibly result in the appearance of bosonic modes only in some special surface orientations. Similar anisotropic electron-phonon interactions have been observed in $MgB_2$ [39], 2H-$NbSe_2$ [40] and hole-doped cuprates [41,42], which suggests a universal picture in understanding these novel superconducting systems, yet a direct link between the anisotropy and the superconductivity is still absent.

Now we move to the second question why these modes are still detected far from the superconducting state, which is seldom observed in other superconducting materials. In conventional superconductors with strong electron-phonon coupling such as Pb [27] and Nb$_3$Sn [28], phonon modes have been verified in tunneling spectra, but their related peaks/dips disappear right above $T_c$. According to the Eliashberg theory, virtual phonons are involved in the electron pairing. Usually, a peak observed at $-\Omega-\Delta$ in negative bias of the $d^2I/dV^2$ curve corresponds to a dip at $\Omega+\Delta$ in positive bias, with $\Omega$ the phonon mode energy [43,44]. The peaks/dips cannot keep in the normal state, as demonstrated by the reduced density of states of Pb (see appendix II). The robust bosonic modes in LTO require us to consider the inelastic process, i.e. an electron injected from the tip to the sample surface emits a real phonon and tunnels into a lower energy state [45,46]. However, it is still interesting why such inelastic process is not commonly observed in other superconducting systems. We remind that there is an evidence for orbital-related state below 50 K in LTO [18]. It is possible that such state modifies the density of states and assists the electron-phonon coupling yet a theoretical model for such speculation is required. Finally, we point out that charge inhomogeneity [47] and spin correlations [42] have been correlated with the anisotropic electron-phonon coupling, which may have resemblance to the orbital-related state effect in LTO.

In conclusion, we present a thorough study on tunneling spectra of high quality LTO thin films deposited on the [111]-, [110]- and [001]-oriented substrates. Several bosonic excitations are identified in the [111]- and [110]-oriented samples except in the [001]-oriented ones. These bosonic modes exist in the superconducting state and extend significantly to the normal state. Our results suggest that the anisotropic spectra come from the anisotropic electron-phonon coupling, which is also reported in other unconventional superconductors. These observations stimulate us to think about the link between the anisotropic electron-phonon coupling and the orbital state, which may further shed light on the nature of the unique superconductivity in LTO.

**ACKNOWLEDGMENTS**
We thank Z. Yuan, R. Z. Huang, L. H. Yang, W. H. Wang, Y. F. Yang, F. V. Kusmartsev for fruitful discussions. This work was supported by the National Key Basic Research Program of China (2015CB921000, 2016YFA0300301, 2014CB921002), the National Natural Science Foundation of China (11674374, 11474338, 51522212, 51421002, 51332001), the Key Research Program of Frontier Sciences, CAS (QYZDB-SSW-SLH008) and the Strategic Priority Research Program of the CAS (XDB07020100, XDB07030200). K.L. was supported by the Fundamental Research Funds for the Central Universities, and the Research Funds of Renmin University of China (14XNLQ03).

**APPENDIX I: DFT calculations and phonon modes**
The first-principles electronic structure calculations were carried out with the projector augmented wave (PAW) method [48,49] as implemented in the VASP package [50-52]. The generalized gradient approximation (GGA) of Perdew-Burke-Ernzerhof (PBE) type [53] was used for the exchange-correlation potential. The kinetic energy cutoff of the plane wave basis was set to be 650 eV. An 8×8×8 k-point mesh was employed for the Brillouin zone (BZ) sampling of the primitive cell. The Gaussian smearing with a width of 0.05 eV was adopted for the Fermi surface broadening. In structural optimization, the experimental lattice constants were used and internal atomic positions were allowed to relax until all forces on atoms were smaller than 0.01 eV/Å. After the equilibrium structures were obtained, the frequencies and displacement patterns of the phonon modes at BZ center were calculated by the dynamical matrix method. The atomic displacements due to the phonon vibrational excitations were set up according to the method of Ref. [54,55]. For the calculations on films, we used simplified lattice distortions with the nonequivalent lattice constants in and out of the *xy* plane for the [001]-oriented film and with the reduced angles (< 60°) between the lattice vectors for the [111]-oriented film.

TAB. A1. Calculated phonon spectra for $LiTi_2O_4$. The frequencies of the $E_g$, $T_{1u}(4)$, $E_u(2)$ and $T_{2u}(2)$ modes are less than 42 meV, which are candidates for the bosonic modes observed in the [110] and [111] films. These energies are close to the frequencies of certain phonons measured by Neutron inelastic scattering [37].

| Mode | Frequencies(cm$^{-1}$) | E(meV) | Neutron results |
|---|---|---|---|
| $A_{1g}$ | 622.4 | 77.2 | |
| $E_g$ | 331.3 | 41.1 | 44 |
| $T_{2g}(1)$ | 532.1 | 66.0 | |
| $T_{2g}(2)$ | 472.7 | 58.6 | |
| $T_{2g}(3)$ | 422.4 | 52.4 | |
| $T_{1u}(1)$ | 561.8 | 70.0 | |
| $T_{1u}(2)$ | 396.0 | 49.1 | |
| $T_{1u}(3)$ | 376.6 | 46.7 | |
| $T_{1u}(4)$ | 279.7 | 34.7 | 34.1 |
| $A_{2u}(1)$ | 592.9 | 73.5 | |
| $A_{2u}(2)$ | 497.3 | 61.7 | |
| $E_u(1)$ | 467.5 | 58.01 | |
| $E_u(2)$ | 203.8 | 25.3 | 26.4 |
| $T_{1g}$ | 385.2 | 47.8 | |
| $T_{2u}(1)$ | 342.4 | 42.5 | |
| $T_{2u}(2)$ | 116.6 | 14.5 | 15.4 |

## APPENDIX II: The origination of the bosonic modes in LTO

To distinguish whether the bosonic modes observed in our experiments origin from virtue phonon process or inelastic process, we compared the behavior of the reduced density of states between Pb and LTO by suppressing the superconductivity. According to Eliashberg-Nambu theory [43,44], the energy gap equations can be expressed as follows:

$$\xi(\omega) = [1 - Z(\omega)]\omega = \int_{\Delta_0}^{\infty} d\omega' Re\left[\frac{\omega'}{(\omega'^2 - \Delta'^2)^{1/2}}\right]$$

$$\times \int d\omega_q \alpha^2(\omega_q) F(\omega_q) [D_q(\omega' + \omega) - D_q(\omega' - \omega)] , \qquad (1)$$

$$\varphi(\omega) = \int_{\Delta_0}^{\omega_c} d\omega' Re\left[\frac{\Delta'}{(\omega'^2 - \Delta'^2)^{1/2}}\right]$$

$$\times \{\int d\omega_q \alpha^2(\omega_q) F(\omega_q) [D_q(\omega' + \omega) - D_q(\omega' - \omega)] - U_c\} , \qquad (2)$$

where $Z(\omega)$ is the normal self-energy, $\varphi(\omega)$ is the pairing self-energy, $\alpha^2(\omega_q)F(\omega_q)$ is effective phonon spectra for phonons of energy $\omega_q$, $U_c$ is Coulomb pseudopotential, $\Delta_0$ is superconducting energy gap, $\omega_c$ is cut off phonon frequency, $D_q(\omega) = 1/(\omega + \omega_q - i\eta^+)$, $\Delta(\omega) = \varphi(\omega)/Z(\omega)$ and $\Delta_0 = \Delta(\Delta_0)$.

In strong coupling theory, the reduced electronic density of states in the superconductor is given by

$$\frac{N_S(\omega)}{N(0)} = Re\left\{\frac{|\omega|}{[\omega^2 - \Delta^2(\omega)]^{1/2}}\right\} , \qquad (3)$$

where $N(0)$ is the electronic density of states at the Fermi surface unrenormalized by the electron-phonon interaction. The final density of states can be obtained by an iteration arithmetic. For the first circulation, the energy gap is set as

$$\Delta(\omega) = \begin{cases} \Delta_0, \omega \leq \omega_D \\ 0, \omega > \omega_D \end{cases} , \qquad (4)$$

where $\omega_D$ is Debye frequency. We need adjust the value of $U_c$ to keep $\Delta_0 = \Delta(\Delta_0)$ in each circulation until $\Delta(\omega)$ converged.

The gap dependence of density of states in Pb is calculated with initial parameters as follows: $\Delta_0 = 1.42\ meV$, $U_c = 0.097$, $\omega_c = \omega_D = 32\ meV$, $\eta^+ = 0.1\ meV$. It is clearly seen that the humps shift to low energy and disappear gradually with decreasing the energy gap as shown in FIG. A1(a).

In tunneling experiments, the normalized differential conductance is proportional to reduced density of states of samples, which can be expressed as follows:

$$\text{DOS}_{eff} = \frac{(dI/dV)_S}{(dI/dV)_N} , \qquad (5)$$

where *S* stands for superconducting state and *N* stands for normal state. We choose the field dependence of tunneling spectra data at 2.5 K in [111]-oriented film with $H_{c2}$ = 11 T so that the thermal broaden effect can be ignored. The data are normalized with the curve at 12 T (see FIG. A1(b)). The observed humps almost disappear after normalization, which are quite different from Pb. Therefore, the humps observed in [110]- and [111]-oriented films mainly origin from inelastic process rather than virtual phonon process.

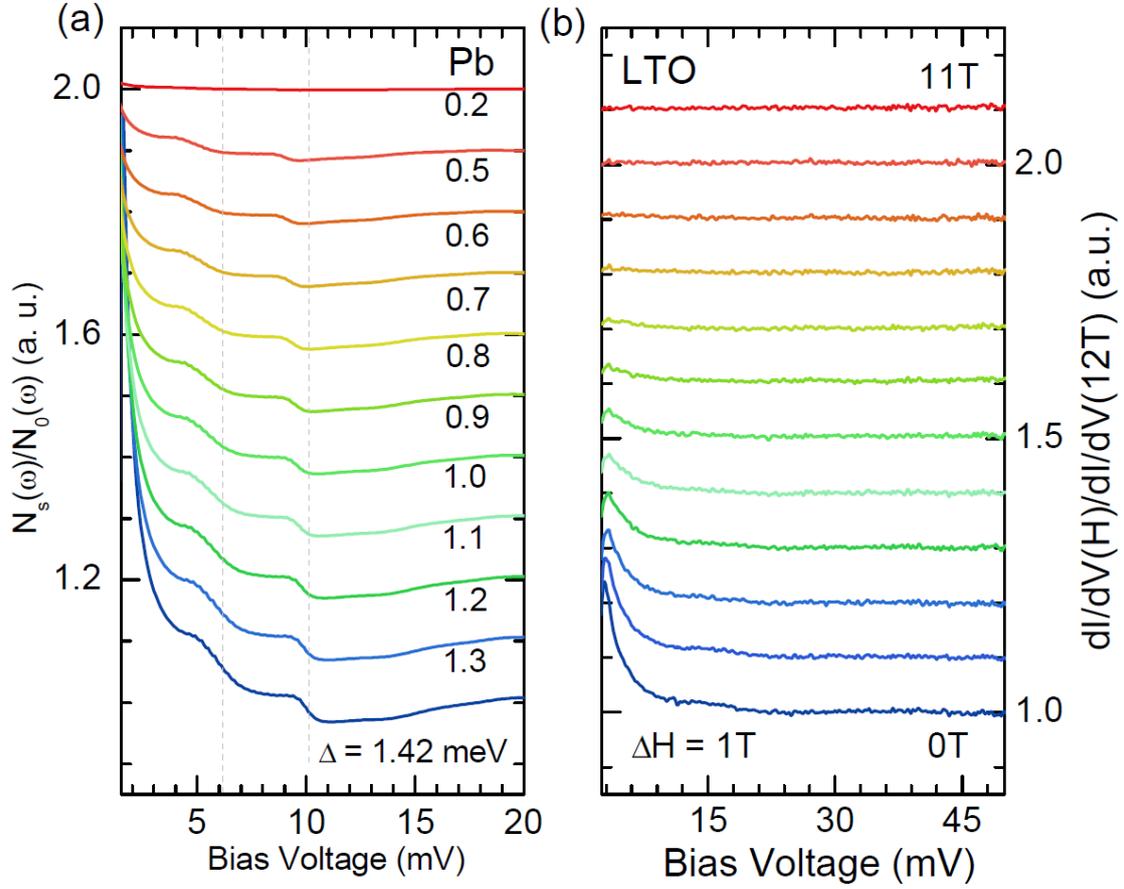

FIG. A1. The comparison of density of states between Pb and LTO. (a) Gap dependence of density of states in Pb calculated by Eliashberg theory. The hump features disappear gradually with decreasing the gap. The dash gray lines are used for guiding the eyes. It can be clearly seen that the hump features shift to low energy with decreasing the gap. (b) Field dependence of normalized tunneling spectra in the [111] samples.

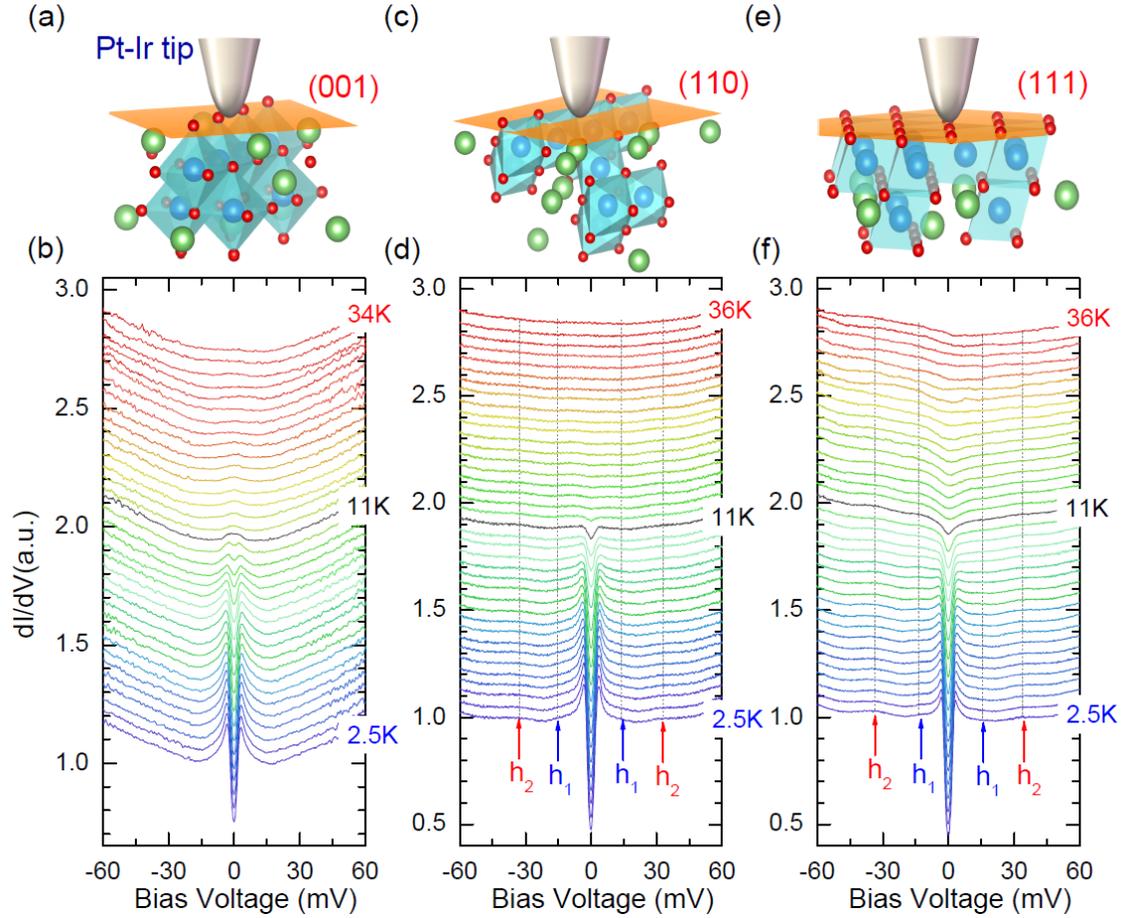

FIG. 1. Temperature dependence of tunneling spectra for spinel superconductor LiTi$_2$O$_4$ thin films grown on [001]-, [110]- and [111]- oriented MgAl$_2$O$_4$ substrates. (a), (c), (e) The crystal plane (001), (110) and (111) of LTO are shown. The red balls stand for O atoms. The green balls stand for Li atoms. The Ti atoms are located at the octahedral sites of oxygen atoms (blue balls). (b), (d), (f) Differential conductance versus bias voltage for the [001]-, [110]- and [111]-oriented LTO thin films from superconducting state to normal state with increasing temperature, respectively. $T \in [2.5\ \text{K}, 10\ \text{K}]$ with $\Delta T = 0.5$ K; $T \in [10\ \text{K}, 12\ \text{K}]$ with $\Delta T = 0.25$ K; $T \in [12\ \text{K}, 14\ \text{K}]$ with $\Delta T = 1$ K; $T \in [14\ \text{K}, 34\ \text{K}\ (36\ \text{K})]$ with $\Delta T = 2$ K.

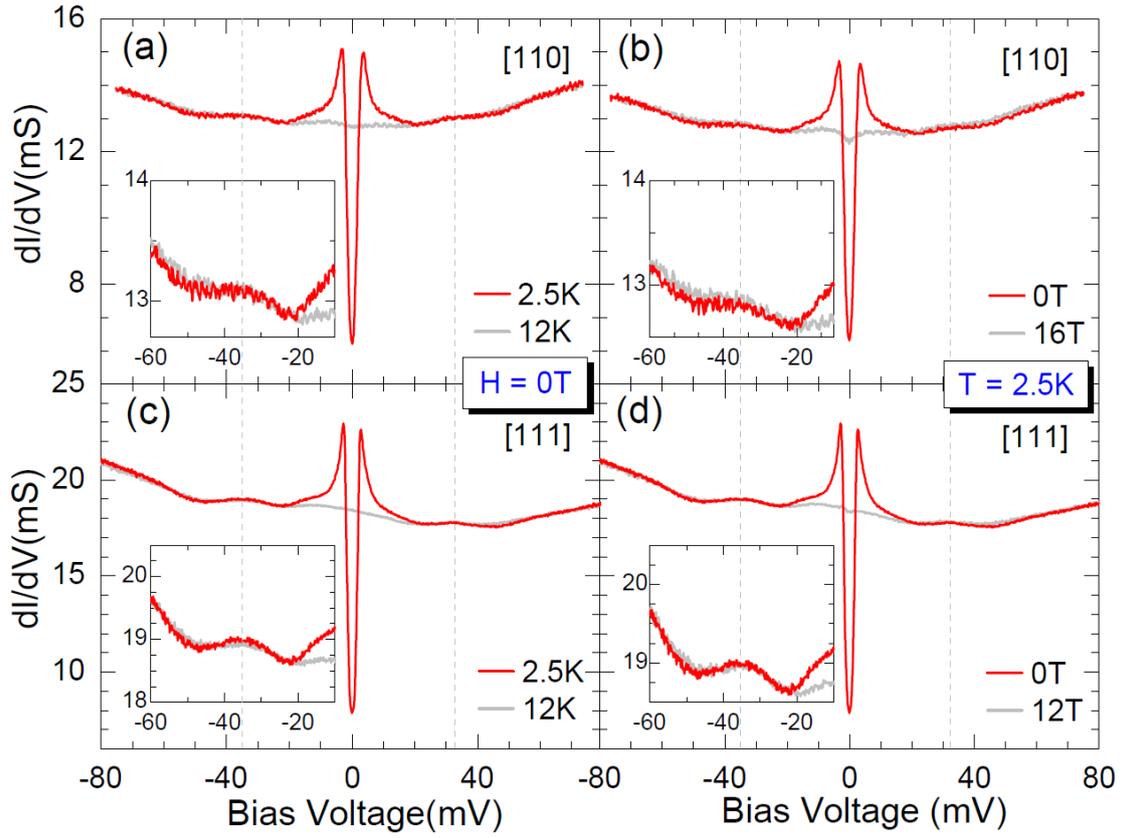

FIG. 2. The comparison of tunneling spectra between the superconducting state and normal state for both [110]- and [111]-oriented samples.(a)~(d), Differential conductance versus bias voltage for (a) [110]-oriented film, $T$ = 2.5 K and 12 K, $H$ = 0 T; (b) [110]-oriented film, $T$ = 2.5 K, $H$ = 0 T and 16 T; (c) [111]-oriented film, $T$ = 2.5 K and 12 K, $H$ = 0 T; (d) [111]-oriented film, $T$ = 2.5 K, $H$ = 0 T and 12 T. The detailed data at negative bias are shown in the inset.

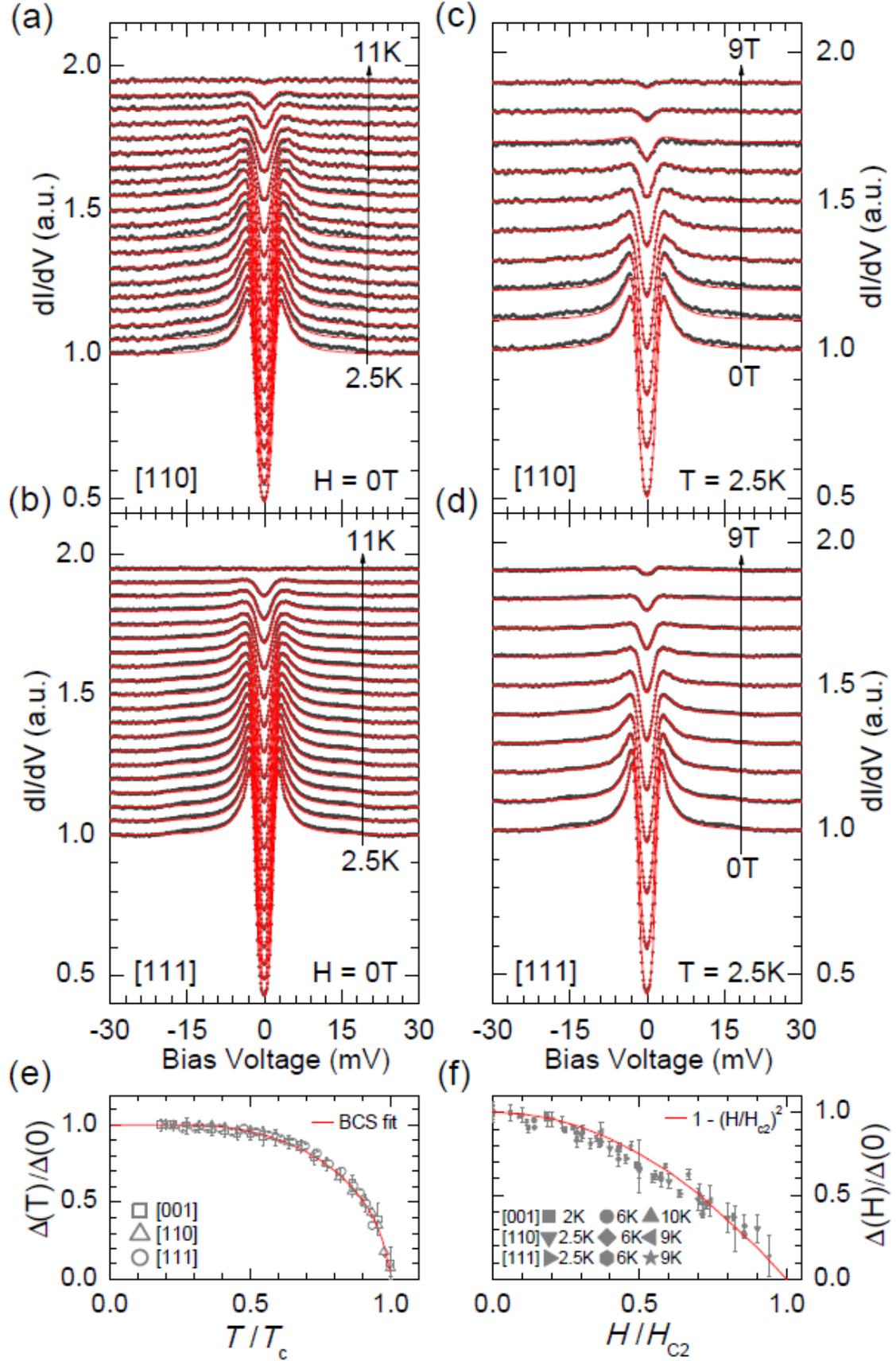

FIG. 3. Fitting results of temperature- and field-dependent tunneling spectra for [110]- and [111]-oriented samples by a modified BTK model. (a), (b) Normalized differential

conductance versus bias voltage for the [110] and [111] films from 2.5 K to 10 K with $\Delta T$ = 0.5 K and from 10 K to 11 K with $\Delta T$ = 0.25 K. Experimental data (black circles) are fitted with a modified BTK model with a constant broadening Γ (red lines). The data are shifted vertically above 2.5 K. (c), (d) Normalized differential conductance versus bias voltage for the (c), [110] and (d), [111] films from 0 T to 9 T with $\Delta H$ = 1 T. Experimental data (black circles) are fitted with an increasing Γ (red lines). (e) Temperature dependence of normalized energy gap $\Delta(T) / \Delta(0)$. The energy gap can be fitted well with BCS theory in all films. $2\Delta/k_\text{B}T_\text{c} \sim 4$ is obtained, indicating a medium-coupling BCS-like superconductor. (f) Normalized energy gap $\Delta(H)/\Delta(0)$ versus normalized field $H/H_{c2}$ in different films, which can be scaled with $1-(H/H_{c2})^2$ for different temperatures.

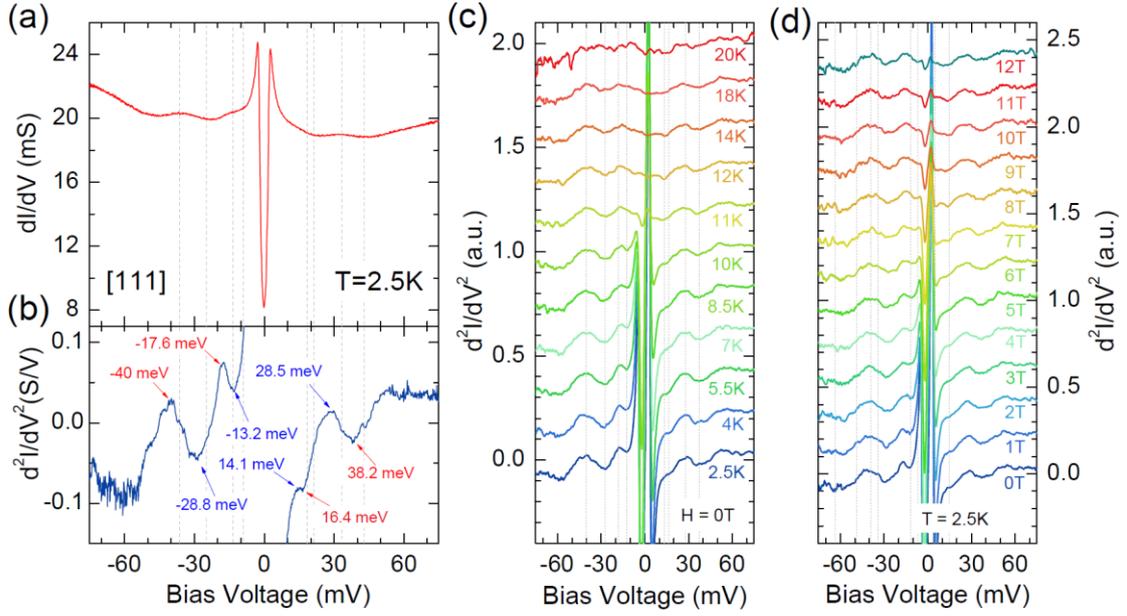

FIG. 4. Temperature- and field-dependent $d^2I/dV^2$ of the [111]-oriented film. (a) The differential conductance versus bias voltage of the [111] film at 2.5 K. Two clear humps can be seen at both negative and positive bias. (b) $d^2I/dV^2$ calculated from the data presented in (a). The bosonic modes are located at the peaks and dips at both negative bias and positive bias with error bars less than ~1 meV. (c), (d) Temperature and field dependence of $d^2I/dV^2$ for the [111] film.

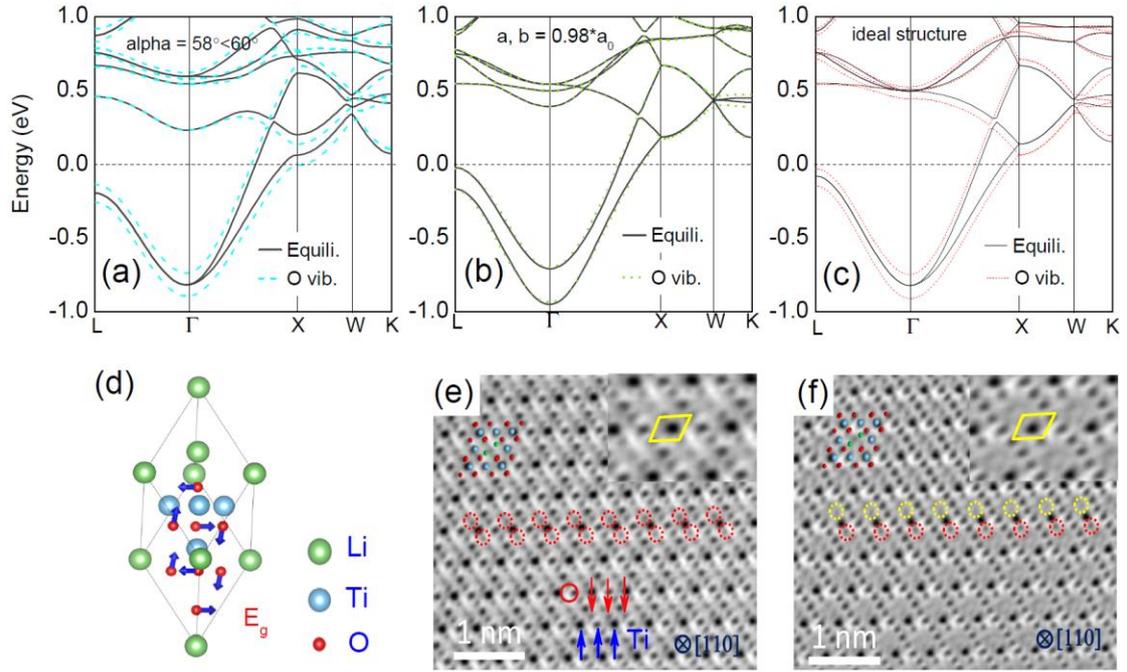

FIG. 5. The origin of anisotropic bosonic modes. Band structures of the LTO films with supposed lattice distortions for (a) the [111]-oriented film with reduced angles (58°) between the lattice vectors of primitive cell and (b) the [001]-oriented film with nonequivalent in-plane (*xy* plane) and out-of-plane (*z* axis) lattice constants. (a), (b) The band structures without and with O atom vibrations according to the $E_g$ (~ 41.1 meV) mode in (d) are denoted by solid and dot/dash lines, respectively. (c) Band structures of ideal LTO structure without (solid lines) and with (short dash lines) O atom vibrations. (d) Atomic displacement patterns of the $E_g$ phonon mode. (e) Typical ABF image of LTO lattice. (f) The representative area showing ordered oxygen vacancy structure. Here, oxygen sites are marked by red dashed circles and the oxygen vacancy sites are marked by yellow dashed circles. The Ti-O octahedron are marked with the same size of yellow quadrilaterals in the insets of (e) and (f) respectively. The octahedron in the ordered oxygen vacancy area cannot match well with the quadrilateral, which indicates a nonnegligible distortion in LTO system.